\documentclass[twocolumn,showpacs,amsmath,amssymb]{revtex4}
\usepackage{graphicx}% Include figure files
\usepackage{dcolumn}% Align table columns on decimal point
\usepackage{bm}% bold math
\usepackage{amssymb}
\usepackage{amsmath}
\usepackage{latexsym}
\def\eg{{\it e.g. }}
\def\etal{{\it et al. }}
\begin{document}

%\title{Heat Conduction in a Quasi One-Dimensional Chaotic Channel}%
\title{The Role of Chaos in One-Dimensional Heat Conductivity}%

\author{Jun-Wen Mao$^{1,2}$}
\author{You-Quan Li$^1$}
\author{Yong-Yun Ji$^1$}
\affiliation{$^1$Zhejiang Institute of Modern Physics,Zhejiang
University, Hangzhou 310027, P.R.China.\\
$^2$Department of Physics, Huzhou University, Huzhou 313000, P.R.
China.}

\date{\today}

\begin{abstract}

We investigate the heat conduction in a quasi 1-D gas model with
various degree of chaos. Our calculations indicate that the heat
conductivity $\kappa$ is independent of system size when the chaos
of the channel is strong enough. The different diffusion behaviors
for the cases of chaotic and non-chaotic channels are also
studied. The numerical results of divergent exponent $\alpha$ of
heat conduction and diffusion exponent $\beta$ are in consistent
with the formula $\alpha=2-2/\beta$. We explore the temperature
profiles numerically and analytically, which show that the
temperature jump is primarily attributed to superdiffusion for
both non-chaotic and chaotic cases, and for the latter case of
superdiffusion the finite-size affects the value of $\beta$
remarkably.

\end{abstract}

% 44.10.+i Heat conduction (see also 66.60.+a and 66.70.+f in transport properties of condensed matter)
% 05.45.-a Nonlinear dynamics and nonlinear dynamical systems (see also section 45 Classical mechanics of discrete systems)
% 05.70.Ln Nonequilibrium and irreversible thermodynamics (see also 82.40.Bj Oscillations, chaos, and bifurcations in physical chemistry and chemical physics)
% 66.70.+f Nonelectronic thermal conduction and heat-pulse propagation in solids; thermal waves (for thermal conduction in metals and alloys, see 72.15.Cz and 72.15.Eb)

\maketitle

\section{Introduction}\label{1}

The low dimensional microscopic dynamics of heat conduction has
been an attractive question since early year last century. Much
more attention has been payed to this problem in the last two
decades due to the dramatic achievement in the application of
miniaturized devices \cite{Forsman,Taillefer,Tighe,Hone,Kim,Zhang}
which can be described by 1D or 2D models. More and more numerical
calculations are focused on the minimal requirements for a
dynamical model where Fourier's law holds or not
\cite{Lepri_1,Casati,Posch,Kaburaki,Lepri_2,Alonso_1,Moasterio,Li_2,Li_3,Alonso_2}.
A convergent heat conductivity was shown in ding-a-ling model
\cite{Casati,Posch} which is chaotic. The studies on Lorentz gas
model \cite{Alonso_1,Moasterio} (the circular scatters are
periodically placed in the channel) of which the Lyapunov exponent
is nonzero gave a finite heat conductivity which fulfils the
Fourier law explicitly. Hence, chaos was ever regarded as an
indispensable factor to normal heat conduction. Whereas, the FPU
model \cite{Kaburaki,Lepri_2} indicated that the chaotic behavior
is not sufficient to arrive at normal heat conduction. Recently, a
series of billiard gas models \cite{Li_2,Li_3,Alonso_1,Alonso_2}
were devoted to explore the normal heat conduction of quasi 1D
channels with zero Lyapunov exponent. However, the role of chaos
in heat conduction has not been well understood. Additionally, the
exponential stability and instability frequently coexist in the
scatters of real system. Thus the model with various degree of
chaos deserves further investigation from the microscopic point of
view, and it will be also interesting to explore non-equilibrium
stationary states and to determine the steady temperature field.

In this paper, we focus on the quasi one-dimensional gas model
which is closer to the real system. The scatters in our model are
the isosceles right triangle with a segment of circle substituting
for the right angle. In this case the edges of scatters are the
combination of line and a quarter of circle. Such a channel is of
chaos, which indicates exponential instability of microscopic
dynamics. Our paper is organized as follows. In section \ref{2},
we introduce the model and investigate the degree of chaos for
various channels with different arc-radius and channel height. In
section \ref{3}, we study the heat transport behavior and the
corresponding diffusive behavior by changing the radius of top arc
and channel height. In section \ref{4}, we investigate the
non-equilibrium stationary state and determine the steady
temperature field numerically. We also analyze the dependence on
diffusion exponent $\beta$ and system size $N$ of temperature
profile theoretically. In section \ref{5}, we discuss the relation
between our work and others and summarize our main conclusions.

\section{the model}\label{2}

\begin{figure}[h]
\includegraphics[width=0.46\textwidth]{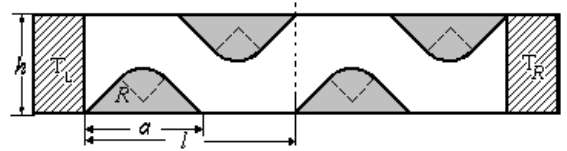}%
\caption{\label{fig:systematic} The channel with $N$ replicate
cells. Here, $l=2.2$, $a=1.2$, $h$ changes from $1.0$ to $0.27$,
and $R$ from $0$ to $0.848528$ to ensure a quarter of circle
always.}
\end{figure}

We consider a billiard gas channel with two parallel walls and a
series of scatters. The channel consists of $N$ replicate cells of
length $l$ and height $h$, and each cell is placed with two
scatters as shown in Fig. \ref{fig:systematic}. The scatter's
geometry is an isosceles right triangle of hypotenuse $a$ whose
vertex angle is replaced by a segment of circle with radius $R$
which is tangential to the two sides of the triangle. At the two
ends of the channel are two heat baths with temperature $T_{L}$
and $T_{R}$. Noninteracting particles coming from these heat baths
are scattered by the walls and the straight lines as well as the
arcs of the scatters in the channel.

%fig1

\begin{figure}[h]
\includegraphics[1.5cm,1cm][8cm,12cm]{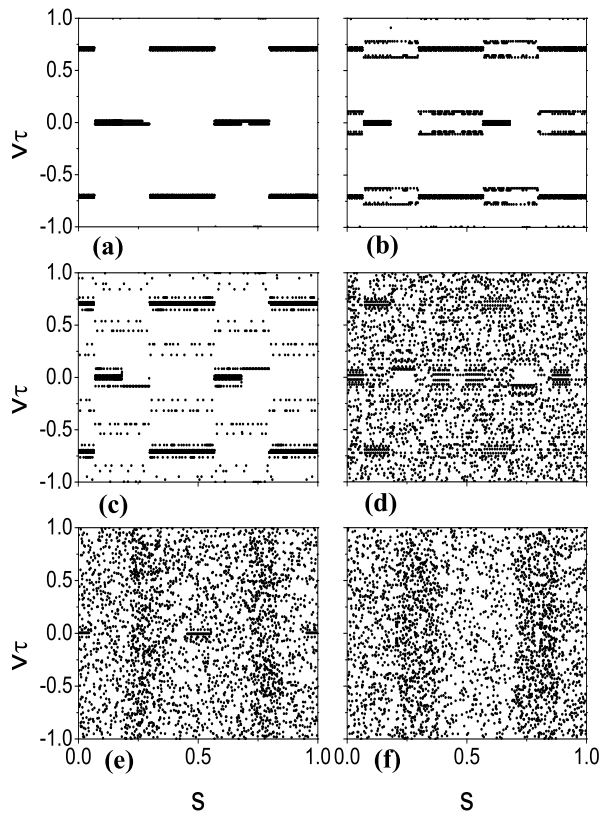}
\caption{\label{fig:sos} Poincare surface-of-section of the
billiard problem. The billiard starts with an incident angle $0.8$
and unit velocity. (a) $R=0$, $h=1.0$; (b) $R=0.001$, $h=1.0$; (c)
$R=0.015$, $h=1.0$; (d) $R=0.1$, $h=1.0$; (e) $R=0.848528$,
$h=1.0$; (f) $R=0.848528$, $h=0.27$. }
\end{figure}

For such a channel, the degree of chaos can be characterized
qualitatively by Poincare surface of section (SOS) \cite{Vega}.
Suppose we take out one unit cell from the channel and close the
two ends by straight walls. Then the problem becomes a billiard
problem. A particle moves within the cell and makes elastic
collision with the mirror-like boundary. We investigate the
surfaces of section ($s$, $v_\tau$) under different initial
conditions. $s$ is the length along the billiard boundary from the
collision point to the reference point. $v_\tau$ is the tangential
component of velocity with respect to the boundary at that point.
The filling behavior of phase space shown in Fig. \ref{fig:sos}
indicates the degree of chaos. In case I, it's non chaotic. The
surface of section is regular and periodic as shown in Fig.
\ref{fig:sos}(a). As the radius $R$ is increased from (a) to (e),
the motion becomes more complex and the map becomes dense with
points except some regular islands. In Fig. \ref{fig:sos}(f), the
regular parts disappear which indicates strong chaos.

%fig2

\section{the heat transport and diffusion behavior}\label{3}

To study the heat conduction of the model, the heat flux is
investigated firstly. In calculating the heat flux, we follow Ref.
\cite{Alonso_1}. For simplicity, the particles from the two heat
baths are supposed to have definite velocities $\sqrt{2T_{L}}$ and
$\sqrt{2T_{R}}$ respectively \cite{Li_2}. We consider one particle
colliding with a heat bath during a period of simulating time. The
energy exchange $(\Delta$E$)_{j}$ at the $j$th collision with the
heat bath is defined as
\begin{equation}
(\Delta E)_{j}=E_{h}-E_{p},
\end{equation}
where $E_{h}$ denotes for the energy of particle taken from the
bath and $E_{p}$ for that carried in the channel. For $M$
collisions between the particle and the bath wall during the
simulation time $t$, the heat flux is given by
\begin{equation}
J_{1} (N)=\frac{\sum_{j=1}^M(\Delta E)_{j}}{t}.
 \label{eq:flux}
\end{equation}

As there is one heat carrier in each cell and the channel has $N$
replicas, there are $N$ particles in the whole channel. Summing
over the heat flux of $N$ heat carriers, we have $J_{N} (N)=N
J_{1} (N)$. Meanwhile, the Fourier's law reads
\begin{equation}
J_{N} (N)=-\kappa\frac{d T}{d x}=\kappa\frac{T_{L}-T_{R}}{Nl},
\label{eq:fourier}
\end{equation}
where $\kappa$ refers to the heat conductivity which is determined
by Eqs. (\ref{eq:flux}) and (\ref{eq:fourier}),
\begin{equation}
\kappa\thicksim N^{2}J_{1} (N).
\end{equation}

\begin{figure}[h]
\includegraphics[2cm,1cm][8cm,6cm]{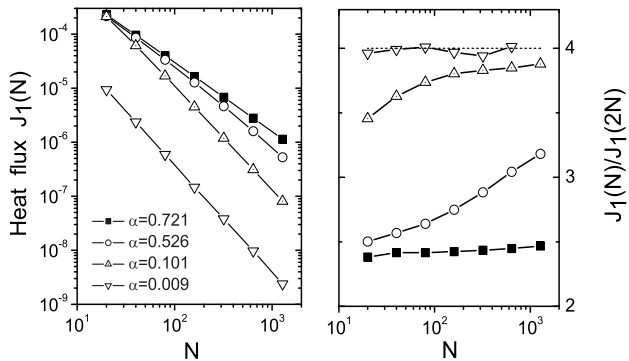}
\caption{\label{fig:heatflow} The heat flux of a single particle
versus system size($N=20$, $40$, $80$, $160$, $320$, $640$ and
$1280$) with the divergence exponent of heat conductivity
$\alpha=0.721$, $0.526$, $0.101$, $0.009$ for four typical cases
respectively (left panel). The ratio of heat flux
$J_{1}(N)/J_{1}(2N)$ for different system sizes (right panel). }
\end{figure}

We consider various cases by changing the radius $R$ of the top
arc of the scatters to investigate their effects on heat
conduction. The heat flux of a single particle versus system size
shown in Fig. \ref{fig:heatflow} are four typical cases. Namely,
case I: the $ {}_{\blacksquare}$ studies for $R=0,\,h=1.0$; case
I\!I: the $\circ$ for $R=0.001,\,h=1.0$; case I\!I\!I: the
$\vartriangle$ for $R=0.848528,\,h=1.0$, and case I\!V: the
$\triangledown$ for $R=0.848528,\,h=0.27$, respectively. The total
cell numbers are chosen as $N=20$, $40$, $80$, $160$, $320$, $640$
and $1280$ respectively. After a sufficient long period of
simulation time, the heat flux approaches to a constant value.
Clearly, the value of heat flux decreases with increasing $R$ for
the same size. Remarkably, there is $20$ times difference of heat
flux between case I\!I\!I and case I\!V, which indicates that
smaller height suppresses the heat flux greatly. Thus, it appears
that the value of heat flux can be adjusted in this way in
designing heat-control devices. Furthermore, our calculations show
that the heat flux dependence on $N$ exhibits faint non-linearity
although the curve looks linear for all cases except case I\!V in
the log-log scale .

%fig3

In order to observe the deviation from the line, which arises from
the finite-size effect, we calculate the ratio of heat flux versus
system size for various radius. The data for the aforementioned
four cases are plotted in the right panel of Fig.
\ref{fig:heatflow}, from which one can see that both the
increasing of system size and of the arc radius bring the ratio an
upward tendency to the value $4$ which ensures the Fourier law. In
case I where $R=0$, there is only a slight increase for the ratio
around $2.4$. Whereas, the ratio rises drastically along with the
increasing of systems size even if the radius $R$ is merely
$0.001$ (the case I\!I). When $R=0.848528$ (the case I\!I\!I), the
scatters become a full segment of quarter circle. The ratio also
rises drastically at first and gradually after $N>160$ in this
case. In both cases I\!I and I\!I\!I, it seemly approaches to
distinct asymptotic values which are all different from that for
normal conduction. This implies that smaller degree of chaos is
insufficient to bring about a normal heat conduction although the
increasingly chaotic degree makes the divergent exponent of heat
conduction smaller. In case I\!V, we maintain the scatters at
radius $R=0.848528$ and reduce the height $h$ from $1.0$ to
$0.27$. In this strongly chaotic case, the ratio fluctuates around
the value $4$ (dotted line) which means that Fourier law is
obeyed.

It is known that the normal heat conduction happens when
$\alpha=0$, which indicates the heat conductivity is independent
of system size, and the anomalous heat conduction corresponds to
the case of $\alpha>0$. The heat conductivity $\kappa$ we
calculated can be given by $\kappa\thicksim N^{\alpha}$ with
$\alpha \gtrsim 0$ despite the heat-flux ratio has a different
increase in asymptotic value for all cases (except case I\!V).

\begin{figure}[h]
\includegraphics[width=0.52\textwidth]{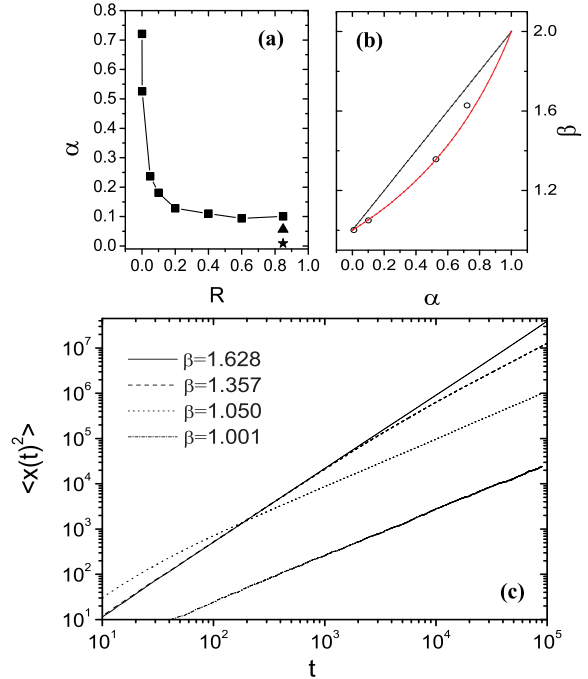}
\caption{\label{fig:alpha} (color on line) (a) Conductivity
divergence exponent $\alpha$ versus circular radius $R$. The
${}_{^\blacksquare}$, refers to the magnitude of $\alpha$ for
$h=1.0$, $R=0,\,0.001,\,0.05,\,0.1,\,0.2,\,0.4,\,0.6,$ and
$0.848528$; the $\blacktriangle$ for $h=0.5$ and $R=0.848528$; the
$\bigstar$ for $h=0.27$ and $R=0.848528$ has the value of $0.009$.
(b) The relation between $\beta$ and $\alpha$, where the circle is
the numerical result, the red line is of $\alpha=2-2/\beta$
\cite{Li_4} and the dashed line is the result of Ref.
\cite{Denisov}. (c) Log-log plot of mean square displacement
$\langle x(t)^{2}\rangle$ versus time $t$. The curves from top to
bottom on the right correspond to cases I, I\!I, I\!I\!I and I\!V
respectively. The ensemble has $10^5$ particles starting from the
center of the channel at time $t=0$ where $x=0$ with the unit
velocity and random direction.}
\end{figure}

We calculate $\alpha$ at the range of system sizes $N$ from $20$
to $1280$ by averaging over many realizations for various radius
$R$ at fixed channel height $h=1.0$, and the plot of the
dependence of $\alpha$ on $R$ is shown in Fig. \ref{fig:alpha}(a).
One can see that $\alpha$ descends from $0.721$ through $0.526$ to
$0.092$ if $R$ increases from $0$ to $0.848528$ for a fixed height
$h=1.0$. Clearly, the $\alpha$ descends rapidly for small radius
(\eg $R=0.001$ in case I\!I) and slowly for larger ones. This
illustrates that the appearance of arc on the top of the scatter
suppresses the divergent exponent $\alpha$ drastically. If the
channel height $h$ for fixed $R=0.848528$ is changed from $1.0$ to
$0.27$, the $\alpha$ is found to diminish to $0.009$. Therefore
the $\kappa$ appears to be independent of system size and the
Fourier law holds in this case.

%fig4

Since the characteristic of heat transport is found being closely
related to the diffusion
behavior\cite{Alonso_2,Li_2,Li_3,Li_4,Li_5,Denisov}, we
investigate the diffusion property for the above cases
subsequently. For a particle starting at the origin at time $t=0$
and diffusing along $x$ direction, the mean square displacement
$\langle(x(t)-x(0))^{2}\rangle$ characterizes its diffusion
behavior. For normal diffusion, the Einstein relation of
$\langle(x(t)-x(0))^{2}\rangle = Dt$ holds, where $D$ is diffusion
coefficient. If the mean square displacement does not grow
linearly in time, {\it {i.e.}}, $\langle(x(t)-x(0))^{2}\rangle =
Dt^{\beta}$, we refer to anomalous diffusion. Recently, the
connection between anomalous diffusion and corresponding heat
conduction in 1D system was discussed hotly
\cite{Li_4,Li_5,Denisov}. We plot the mean square displacement
versus time $t$ in Fig. \ref{fig:alpha}(c) for the aforementioned
four cases. Note that $10^{5}$ particles were put at the center of
the channel where $x=0$ with unit velocity and random direction in
the simulations. The top solid line and the bottom short-dash-dot
line are precisely straight in the whole simulation period
($t=10^5$), which correspond to the case I (non-chaotic) and case
I\!V (strong chaotic), respectively.  We obtain $\beta=1.628$ for
the case I which corresponds to $\alpha=0.721$, and $\beta=1.001$
to $\alpha=0.009$ for the case I\!V. Beyond these two cases do the
curves keep asymptotically linear at large time $t$ with diffusion
exponent $\beta$ between the values of above two cases. The best
fits of the slope give $\beta=1.357$ which corresponds to
$\alpha=0.526$ for case I\!I and $\beta=1.050$ to $\alpha=0.101$
for case I\!I\!I, respectively. The relation between divergent
exponent $\alpha$ and diffusion exponent $\beta$ fits the relation
of $\alpha=2-2/\beta$ proposed by Li and Wang in Ref. \cite{Li_4},
as is plotted in Fig. \ref{fig:alpha}(b). Whereas Denisov {\it et
al.} presented another connection of $\alpha$ with $\beta$ on the
basis of the L$\acute{e}$vy walk model \cite{Denisov}. More
details about the origin of the discrepancy between above two
relations can be found in Ref. \cite{Li_5}.

%fig5
\begin{figure}[h]
\includegraphics[width=0.54\textwidth]{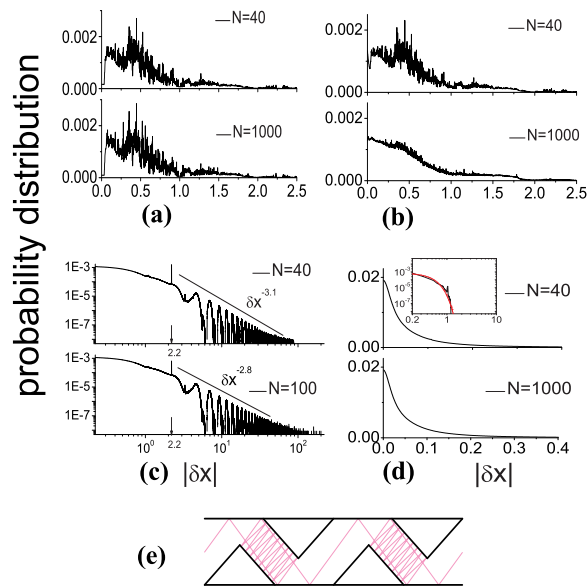}
\caption{\label{fig:MFP} (color on line) (a), (b), (c) and (d) The
PDFs of the flight distance $|\delta x|$ between two consecutive
collisions for above four cases I, I\!I, I\!I\!I and I\!V
respectively. $N$ represents the system size. Note that the
log-log scale is used in (b) and the inset of (d). In the latter
case (d), Gaussian distribution (red dashed-line) is in comparison
to the numerical PDF. (e) The typical trajectory with periodicity
in case of $R=0$ , $h=1.0$.}
\end{figure}

As different diffusion behaviors are likely related to the
trajectory characteristics of the particle propagation, we
investigate the PDF $\psi(|\delta x|)$ of the flight distance
$|\delta x|$ in $x$-direction between two consecutive collisions
with the scatters. After a long time for adequate collisions in
the channel, the PDF for aforementioned fore cases, shown in Fig.
\ref{fig:MFP}(a) to (d) respectively, take on completely different
forms for different cases. In case I, the discrete values of
probability indicate that the trajectories are abundant of
periodicity, which is almost alike for larger system size. The
maximum value of PDF appears when $|\delta x|=0.447$, and the
typical trajectory is plotted in Fig. \ref{fig:MFP}(e) which shows
explicitly that the parallel passage makes the periodical
trajectory possible, and the particles are easier to propagate
along the channel with fewer collisions. It is superdiffusion in
this case. In case I\!I, only smaller system size has the explicit
periodicity. With the system size growing, the periodicity is
destroyed by the collisions with the segment of circle time and
time. The PDF gets smoother in this case. In case I\!I\!I, the
periodicity happens only for large flight distance $|\delta x|$
with very small number of families. Note that the maximum of PDF
is corresponding to the value $|\delta x|$ of $2.2$ which is just
the length of a cell, and the PDF decays in power law. In this
case it requires more collisions and takes more time for the
particles to escape a certain region. Thus the propagation is
suppressed but is still of superdiffusion. The normal diffusion
takes place when the particles are scattered by sufficiently large
density of hyperbolic scatters (case I\!V). Consequently, the
strong chaos presents the trajectory of heat carriers with more
aperiodicity. The PDF takes on its characteristic form which has a
Gaussian tail as shown in the inset of Fig. \ref{fig:MFP}(d).

Thus, the propagation modes are responsible for the diffusion
behavior. The abundance of aperiodicity of trajectory is the
characteristic of chaotic channel and may also play an crucial
role in the normal diffusion. In other words, if the trajectory in
a certain system emerges the aperiodicity due to some other
mechanisms, such as in polygonal billiard gas model
\cite{Alonso_2}, the normal diffusion behaviors may happen.

\section{The calculation of temperature field }\label{4}

\begin{figure}[h]
\includegraphics[width=0.48\textwidth]{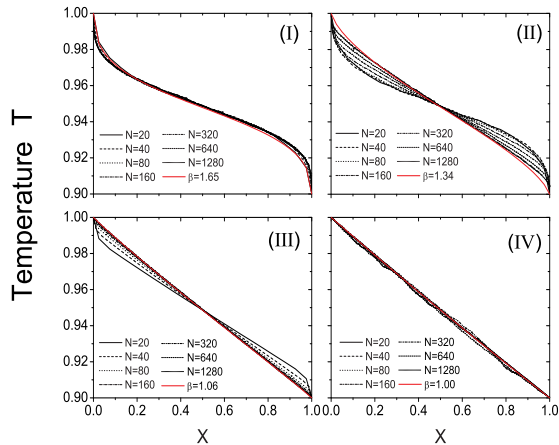}
\caption{\label{fig:temp} (color on line) Numerical results of
temperature profiles for $T_{L}=1.0$, $T_{R}=0.9$ and sizes
$N=20$(solid), $40$(dash), $80$(dot), $160$(dash dot), $320$(dash
dot dot), $640$(short dash) and $1280$(short dot), respectively.
The four panels refer to (I)$R=0,\,h=1.0$; (I\!I)
$R=0.001,\,h=1.0$; (I\!I\!I) $R=0.848528,\,h=1.0$, and (I\!V)
$R=0.848528,\,h=0.27$. The red lines correspond to the best fits
for the numerical temperature profile at $N=1280$ with Eq.
(\ref{eq:temp}), giving the analytical values $\beta$ with $1.65$,
$1.34$, $1.06$ and $1.00$ for above four cases respectively. }
\end{figure}

We calculate the temperature field following the approach proposed
in Ref. \cite{Alonso_1}. The temperature of $i$th cell is defined
by averaging the kinetic energy over all visits into the cell
\begin{equation}
T_{i}=<E_{i}>=\frac{\displaystyle\sum_{j=1}^{m}t_{j}E_{ij}}{\displaystyle\sum_{j=1}^{m}t_{j}},
\end{equation}
where $t_{j}$ denotes for the time spent within the cell in the
$j$th visit, and $m$ for the total number of visit. For
sufficiently large $m$ we expect a steady temperature profile, and
this is indeed verified in our calculations for totally $10^{10}$
visits. The temperature profiles we obtained are plotted in Fig.
\ref{fig:temp}. It is worthwhile to point out that the steady
temperature profiles between non-chaotic and chaotic system are
quite different in thermodynamics limit, which is due to the
different diffusion behaviors as shown in Fig. \ref{fig:alpha}(c).
As case I is non-chaotic and has uniform diffusion exponent
$\beta$, the temperature profiles keep almost the similar shape
for different system sizes. At the two ends of channel there are
large temperature jumps which play an important role in the
Fourier transport and dynamics of the system \cite{Aoki}. These
jumps arise from the boundary heat resistance which usually
appears when there is a heat flux across the interface of the two
adjacent materials. In case I\!I and I\!I\!I which are chaotic and
have asymptotically decreasing $\beta$ versus time $t$, there also
exists the boundary heat resistance. Unlike in case I, the
temperature jump here is smaller and diminishes when the system
size grows. For larger size is there almost no temperature jump
which corresponds to a nearly linear temperature profile. Both the
larger system size $N$ and the arc radius lead to the increase of
chaos degree which is responsible for the decrease of diffusion
exponent $\beta$ ($\geqslant 1$). In case I\!V, which is strong
chaotic, the temperature profiles are almost linear for various
system sizes we considered, corresponding to the normal heat
conduction.

%fig6

We estimate the temperature profiles from the average point of
view. Considering the incident particles from the left heat bath
(where $x=0$) propagating along the x-axis to the right end, we
suppose that a reflecting boundary is placed at the origin of the
x-axis and an absorbing one at the other end. When
$2\geqslant\beta\geqslant 1$, we assume that the mean density
$n_{L}(x)$ of the particles at site $x$ in the steady state is
proportional to $(1-x)^{\gamma}$ with
$\gamma=(2/\beta-1)\beta^{3/2}$, where we set $x=i/N$. Under this
assumption, we have $n_{L}(x)\sim 1-x$ \cite{Alonso_1} when
$\beta=1$ (normal diffusion) and $n_{L}(x)\sim const$ when
$\beta=2$ (ballistic diffusion). The conservation of particle
number requires
\begin{equation}
n_{L}(x)\sim \frac{(1-x)^{\gamma}}{D_{L}},
\end{equation}
where $D_{L}$ are the diffusion coefficient. Likewise, for a
particle propagating from right to left,we have
\begin{equation}
n_{R}(x)\sim \frac{x^{\gamma}}{D_{R}},
\end{equation}

We assume $D_{L}=T_{L}^{\beta/2}$ and $D_{R}=T_{R}^{\beta/2}$.
Thus, if $2\geqslant\beta\geqslant 1$, the temperature is given by
\begin{eqnarray}
&&T(x)=\frac{T_{L}n_{L}(x)+T_{R}n_{R}(x)}{n_{L}(x)+n_{R}(x)}\nonumber\\
&&
=\frac{T_{L}T_{R}^{\beta/2}(1-x)^{\gamma}+T_{L}^{\beta/2}T_{R}x^{\gamma}}
{T_{L}^{\beta/2}x^{\gamma}+T_{R}^{\beta/2}(1-x)^{\gamma}}.
\label{eq:temp}
\end{eqnarray}
where $\gamma=(2/\beta-1)\beta^{3/2}$.

%fig7
\begin{figure}[h]
\includegraphics[width=0.4\textwidth]{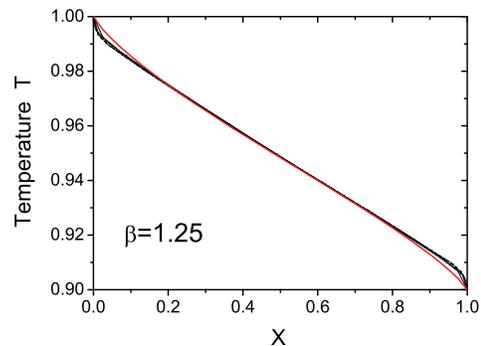}
\caption{\label{fig:comp}(color on line) Numerical results of
temperature profile in comparison to the analytical results. The
numerical temperature profiles with $R=0.848528, N=40$; $R=0.4,
N=40$; $R=0.2, N=40$; $R=0.1, N=80$; $R=0.05, N=160$ at $h=1.0$
and $R=0.848528, N=20$ at $h=0.5$ almost share the same shape. The
red line is the plot of Eq.(\ref{eq:temp}) with $\beta=1.25$. }
\end{figure}

As shown in Figs. \ref{fig:temp} and \ref{fig:comp}, the
analytical results (in red lines) are in good agreement with the
numerical ones for all the cases except those at the two ends of
the channel for superdifusion cases. These deviations are likely
due to the different boundary conditions we used. Furthermore, the
value of $\beta$, obtained by the best fits for the numerical
temperature profile at $N=1280$ with Eq. (\ref{eq:temp}), agrees
with the simulating result greatly for aforementioned four cases.
One can see clearly from Eq.(\ref{eq:temp}) that temperature
profiles are closely related to the diffusion exponent $\beta$,
namely, the case with smaller diffusion exponent tends to have
smaller temperature jump. Accordingly, it is not unexpected that
different chaotic cases may share the same temperature profile if
they have the identical diffusion exponent $\beta$. As shown in
Fig. \ref{fig:comp}, the case with smaller diffusion exponent
requires smaller system size for achieving the same temperature
profile. Moreover, our calculations show that the results of
Eq.(\ref{eq:temp}) are consistent with the numerical ones even in
larger temperature gradient. Thus, the temperature profile is
mostly dependent on the diffusion behavior which is remarkably
affected by the finite-size effect for chaotic cases of
superdiffusion.

\section{discussion and conclusion}\label{5}

In summary, we have investigated the role that the chaos plays in
the heat conduction by billiard gas channel. We have demonstrated
that the degree of dynamical chaos is enhanced by increasing the
arc radius or the system size for chaotic channel, and the mass
and heat transport behavior is significantly related to the degree
of dynamical chaos of a channel. The stronger the chaos is, the
closer to normal transport behaviors the model seems to be.
Furthermore, our numerical results of two exponents $\alpha$ and
$\beta$ for both non-chaotic and chaotic cases when
$\beta\geqslant 1$ satisfies the formula $\alpha=2-2/\beta$
\cite{Li_4}. We also discussed the microscopic dynamics by the PDF
of flight distance in $x$-direction. It seems that aperiodicity of
trajectory plays an important role in diffusion behavior. Finally,
our results showed that the temperature jumps at both ends of the
channel depend mostly on the diffusion property for both
non-chaotic and chaotic channels, and the finite-size effect is
more crucial for chaotic ones.

As is known that the billiard gas model is applicable for
capturing the underlying dynamics of particles without
interaction. It is therefore worthwhile to discuss the relation
between our work and others in this field.

Alonso \etal \cite{Alonso_1} investigated 1D Lorentz gas model
full of periodically distributed half circular scatters. By
defining the heat conductivity and temperature field as
statistical average over time on the hypothesis of local thermal
equilibrium, the Fourier law holds in this case, and a linear
gradient is given for quite small temperature difference . Our
work starts from the same approach but different scatter geometry
is taken into account. Thus it is not surprising that our work has
some overlap with theirs in spirit. However, we pay much more
attention to the role played by different degree of dynamical
chaos in heat conduction. As a result, our intensive calculations
extended the results in  Ref. \cite{Alonso_1} and concluded that
only sufficient strong chaos results in the normal diffusion,
thus, the normal heat transport.

Li \etal \cite{Li_2} presented the dependence of heat conductivity
on system size and the temperature profile in channel with zero
Lyapunov exponent where the right triangle scatters are
periodically distributed. In this case, the exponent stability
leads to abnormal transport behavior. Clearly, their result is the
non-chaotic limit of our model.

\section*{Acknowledgement}

We would like to thank B. Li for providing Ref. \cite{Li_5} prior
to publication and helpful discussion. The work is supported by
NSFC No.10225419 and 90103022.

\end{document}